\documentclass[aps,showpacs,preprint,prb,ONEcolumn]{revtex4}
\usepackage{mathrsfs}
\usepackage{amssymb}
\usepackage{amsmath}
\usepackage{bm}
\usepackage{graphics}
\usepackage{natbib}

\begin{document}
\title{Suppressed conductance in a metallic graphene nano-junction}
\author{Haidong Li}
\author{Lin Wang }
\author{Yisong Zheng }\email[Correspondence author. Email: ]{zys@mail.jlu.edu.cn}
\address{Department of physics, Jilin University, Changchun 130023, China}
\date{\today}

\begin{abstract}
The linear conductance spectrum of a metallic graphene junction
formed by interconnecting two gapless graphene nanoribbons is
calculated. A strong conductance suppression appears in the vicinity
of the Dirac point. We found that such a conductance suppression
arises from the antiresonance effect associated with the edge state
localized at the zigzag-edged shoulder of the junction. The
conductance valley due to the antiresonance is rather robust in the
presence of the impurity and vacancy scattering. And the center of
the conductance valley can be readily tuned by an electric field
exerted on the wider nanoribbon.
\end{abstract}
% \keywords{}
 \maketitle
\bigskip
Graphene, an atomically thin layer of graphite, is regarded as a
perspective base for the post-silicon electronics since its first
experimental realization.\cite{refNovoselov1} Motivated by possible
device applications, the electronic and transport properties of
various graphene nano-structures have been studied both
experimentally and theoretically,
\cite{refD.A.Areshkin,refB.Obradovic,refM.Y.Han,refY.Ouyang,refA.Rycerz,refT.B.Martins,refhod,refL.Brey,refZ.p.Xu,refS.Hong,refT.C.Li}
Among these structures, graphene nanoribbon(GNR) is the basic
element to carry the current. Band structure calculations indicate
that the zigzag-edged graphene nanoribbons are always metallic while
the armchair-edged ones are either metallic or semiconducting,
depending on the width of the armchair GNR.\cite{refL.Brey} A
graphene junction can be formed by interconnecting two semi-infinite
GNRs with different widths. In such a graphene nanostructure, a
traveling carrier is scattered by the junction interface, which
causes a finite junction conductance. Recently, the conductance
spectrum of a graphene metal-semiconductor junctions has been
studied in details.\cite{refT.C.Li,refZ.p.Xu,refS.Hong} It has been
found that the presence of the lattice vacancy can efficiently
enhance the junction conductance, because that a vacancy makes the
coupling between the electron states of the two GNRs at the junction
interface stronger.
\par
\begin{figure}
\centering \scalebox{0.4}{\includegraphics{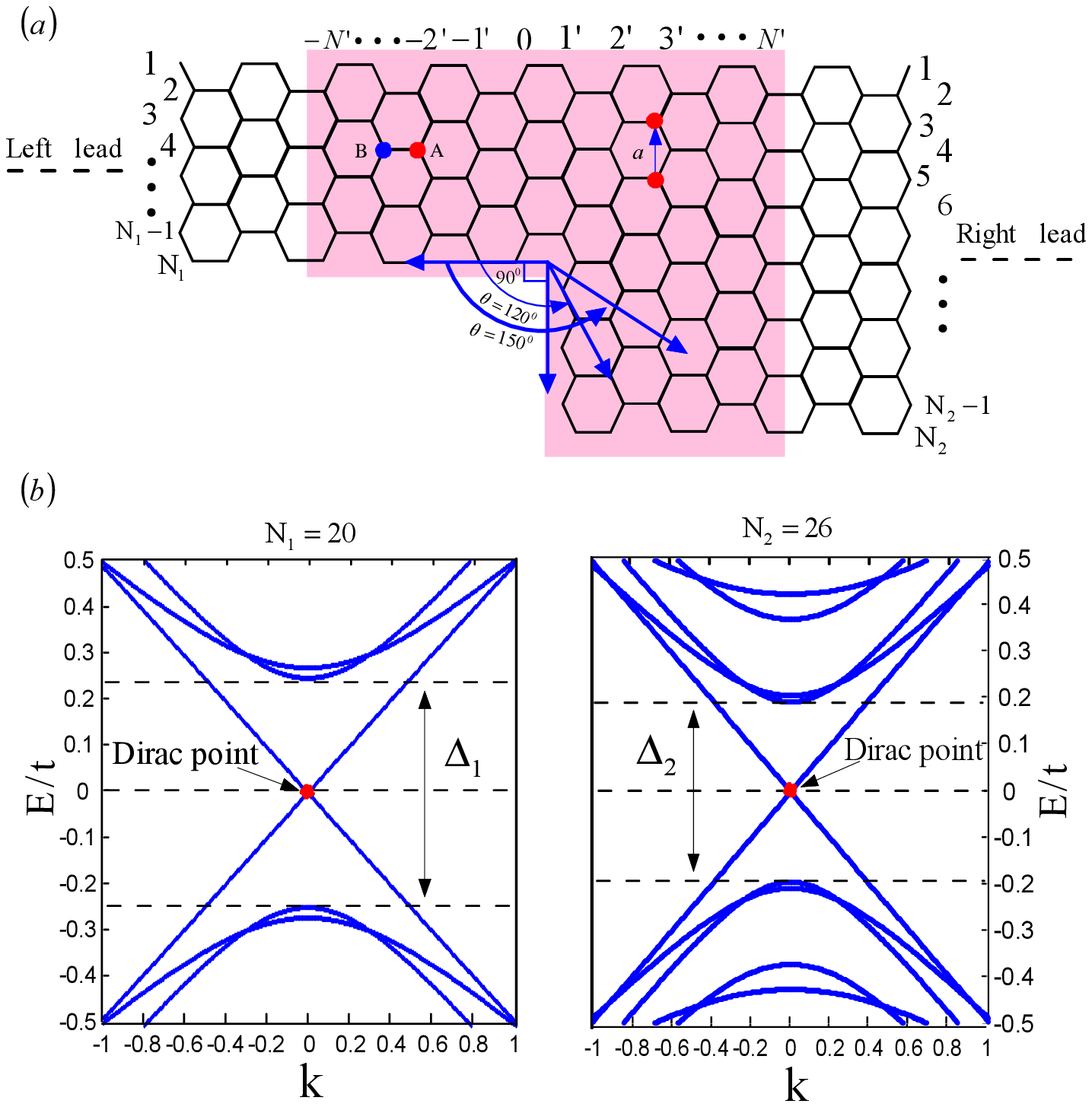}}
\caption{(Color online)(a) Schematic of the metallic graphene
junction. Around the junction interface, a device region(shaded) is
defined where the impurities are randomly distributed. (b) The
subband structures of the two metallic armchair ribbons that
constitue the junction. The ribbon width satisfies $N_{1(2)}=3m-1$
with $m$ being an arbitrary integer. For a charge neutral junction,
the Dirac points of the two ribbons are aligned with each other .
\label{schematic}}
\end{figure}

Apart from the metal-semiconductor junction, a metallic graphene
junction can be constructed by interconnecting two metallic armchair
GNRs with different widths. In this letter, we investigate the
conductance spectrum of such a metallic graphene junction, the
lattice structure of which is depicted in Fig.\ref{schematic}(a).
Unlike a metal-semiconductor junction,\cite{refZ.p.Xu,refS.Hong} the
gapless subband structures of the two metallic armchair nanoribbons
shown in Fig.\ref{schematic}(b) indicates that electronic
transmission through the metallic junction with an arbitrary energy
is formally allowed. In particular, the electronic transmission via
the linear subbands is reflectionless even in the presence of a
long-range scattering potential, due to  pseudospin
conservation.\cite{refM.I.Katsnelson} Therefore, a plausible
anticipation is that the electronic transmission probability near
the Dirac point should be close to unity. However, our calculation
gives the opposite result: the junction conductance at the Dirac
point is equal to zero, and a conductance valley appears around this
point. This means a strong conductance suppression in the vicinity
of the Dirac point. Further analysis demonstrates that the
conductance suppression arises from the antiresonance due to the
existence of an edge state localized in the shoulder region of the
junction.
\par
To calculate the linear conductance of the metallic graphene
junction $\mathcal{G}(E)$ as a function of the incident electron
energy $E$, we adopt the Landauer-B\"utikker formula in the discrete
lattice representation.\cite{refF.M,Datta} It gives
$\mathcal{G}(E)={2e^2\over h}\text{Tr}(\Gamma_1G\Gamma_2G^\dag)$,
where $G=[E+i0^+-H_d-\Sigma_1-\Sigma_2]^{-1}$ is a retarded Green
function, and $H_d$ is the tight-binding Hamiltonian of the device
region in the nearest-neighbor approximation. The contributions of
the two semi-infinite leads are incorporated by the two self-energy
terms $\Sigma_{1(2)}$ which are associated with the coupling
functions $\Gamma_{1(2)}$ by
$\Gamma_{1(2)}=i[\Sigma_{1(2)}-\Sigma_{1(2)}^\dag]$. The two
self-energy terms are obviously the key quantities for calculating
the conductance, which can be evaluated by the recursive
method.\cite{refF.M} In what follows we use the lattice constant $a$
and the hopping energy $t$ between the nearest neighbor atoms as the
units of the length and energy.
\par
The calculated conductance spectra($\mathcal{G}$ vs $E$) for some
typical square junctions($\theta=90^\circ$) are shown in Fig.2(a).
All these conductance spectra exhibit the staircase-like structures,
which can be readily explained by matching of the subband structures
of the two component ribbons as shown in Fig.\ref{schematic}(b).
Another point to note in Fig.2(a) is that the conductance spectrum
shows a notable suppression in the vicinity of the Dirac point when
the difference of the widths of the two GNRs is larger than 3. In
particular, a zero conductance occurs at the Dirac point. However,
from Fig.1(b) we can readily find that the linear subbands of the
two ribbons always match each other to provide an electron
transmission mode, and hence a nonzero conductance at the Dirac
point, in contradiction with the calculated zero conductance. As
shown in Fig.2(b), we find that the conductance suppression near the
Dirac point is tightly associated with the zigzag edge of the
shoulder of the junction. On the contrary, when the edge of shoulder
is of an armchair type($\theta=120^\circ$), the conductance spectrum
no longer shows any suppression. From such a result we infer that
the nature of the conductance suppression is the antiresonance
effect, the detail of which is as follows. In the vicinity of the
Dirac point, only the linear subbands are relevant to the electron
transmission. Hence the two metallic GNRs can be viewed as single
mode quantum wires coupling to each other directly. However, the
zigzag-edged shoulder induces a localized edge state with the
eigen-energy equal to the Dirac energy. Such a localized state
couples to the linear subband of the wider GNR. Consequently, when
the electronic transport is limited to the vicinity of the Dirac
point, the graphene junction is equivalent to the T-shaped quantum
dot structure as shown in Fig.2(c). The linear conductance of such a
model has been extensively studied\cite{refA.
Ueda,refY.Liu,refW.J.Gong} and can be expressed in terms of the
model parameters
\begin{equation}
\mathcal {G}(E)={2e^2\over
h}\frac{2\xi}{(1+\xi)^2}\frac{[E-\varepsilon]^2}
{(E-\varepsilon)^2+[\frac{\Gamma}{2(1+\xi)}]^2},\label{CF}
\end{equation}
where $\Gamma=2\pi\rho_2 \tau^{2}$ and $\xi =\pi^2\rho_1\rho_2v^2 $
with $\rho_{1(2)}$ being the electron density of the states in two
leads. This expression presents a zero conductance at the quantum
dot level $\varepsilon$, which is called the antiresonance effect.
The antiresonance is in fact a result of quantum interference. The
lateral quantum dot introduces new Feynman paths with a phase shift
$\pi$. As a result, the destructive quantum interference occurs
among electron Feynman paths.\cite{refY.Liu,refW.J.Gong} In the
metallic graphene junction, the edge state attached to the
zigzag-edged shoulder of the junction plays a role of laterally
coupled quantum dot, which results in the antiresonance at the Dirac
point.
\par
When the widths of the two GNRs are fixed, the geometry of the
junction can be changed by shifting downwards the narrower ribbon.
In Fig.2(d) the conductance spectra are compared for differently
shaped junctions. We can see that the width of conductance valley
depends on the junction shape sensitively. To be more specific, in
terms of the valley width the spectra are classified into three
groups, each of which appears periodically whenever the narrower
ribbon shifts downwards by three multiples of the lattice constant.
This phenomenon can be explained with the help of the above quantum
dot model. The parameter $v$ in the quantum dot model is a relevant
quantity to the width of the antiresonance valley. From
Fig.\ref{schematic}(a) we can see that $v$ is proportional to the
product of the electron probability amplitudes of the A and B atoms
interconnecting directly at the junction interface. And the
probability amplitudes can be obtained by solving the Dirac equation
\cite{refL.Brey}. In such a way we work out the following relation
$v\propto \sum_{j\in odd}\sin (2j\pi/3)\sin[2(j+n_0)\pi/3]$, where
$j$ is any odd number within the range from 1 to $N_1$, and $n_0$
denotes the displacement of the narrower ribbon with respect to the
upper edge of the wider ribbon. From this relation we can readily
understand the periodic feature of the valley width of the
conductance spectra shown in Fig.2(d).
\par
The antiresonance picture of the conductance suppression in the
metallic graphene junction is further demonstrated by the calculated
spectra of the local density of states (LDOS) at some lattice points
near the junction interface. From Fig.3 we can see that only for the
lattice points at the zigzag edge of the shoulder of
junction($\theta=90^\circ$ and $150^\circ$), the LDOS spectrum
exhibits a very sharp peak at the Dirac point. This indicates the
existence of a localized state in the shoulder region of the
junction. An exception occurs for the junction with a very short
shoulder($N_1=20$ and $N_2=23$ in Fig.3(a)). The LDOS spectrum of
the lattice point at the zigzag-edged shoulder does not show a
notable peak. This result is consistent with the previous
work\cite{refhod}, which argued that a zigzag edge of the width
smaller than three lattice constant can not induce any localized
edged state.
\par
We now proceed on to discuss the influence of the possible
scatterers in an actual graphene junction on the antiresonance
valley. At first, we consider the individual effect of an impurity
appearing at distinct positions. In the numerical calculation, an
impurity is simulated by the deviation of the on-site energy of an
specific lattice point where the impurity appears. The calculated
conductance spectra with an individual impurity positioned at
different lattice points are compared in Fig.4(a). We can see that
only when the impurity appears in the shoulder region, the variation
of the conductance spectrum is notable. But the effect of such an
impurity is not to destroy the antiresonance at the Dirac point.
Instead it causes another conductance zero near the Dirac point.
This result implies that the edge state is intricately affected by
an impurity in the shoulder region. On the contrary, the impurity at
other lattice points can not modify the antiresonance valley notably
since it is irrelevant to the edge state. Fig.4(b) shows the
conductance spectrum in the presence of many impurities distributed
randomly in the device region with fluctuating strengths. We can see
that the antiresonance is rather robust even if the impurity
concentration and strength are nontrivially large. Fig.4(c) shows
the effect of vacancies positioned uniformly at the zigzag edge of
the shoulder of a square junction. A vacancy is simulated in the
numerical calculation by simply cutting off a carbon atom of type A
at the edge of the shoulder. If we use the notations $M_1$ and $M_2$
to denote the atom number of type A belonging to the edge of the
shoulder and the number of the vacancy in this edge respectively,
the ratio $r=M_2/M_1$ can be viewed as the concentration of vacancy.
From Fig.4(c) we can see only when the concentration of the
vacancies is about $r=1/3$, the zero conductance at the Dirac point
can be completely eliminated. Higher or lower concentrations of
vacancies can not destroy the conductance valley around the Dirac
point. Our calculation also indicates that such a conclusion is
independent of the size of the shoulder. This result implies the
complicated effect of the edge defect on the edge state. Such an
interesting topic is left for our study in the future. Finally, we
can apply a step-like potential in the right hand side of the
junction to tune the position of antiresonance, which can be
simulated by shifting the on-site energy of all the lattice points
of the wider GNR. The result is shown in Fig.4(d). We can see that
the step-like potential simply shifts the antiresonance point,
without drastically altering the lineshape of the conductance
spectrum. What is noteworthy is that the conductance at the Dirac
point can be easily tuned from zero to unity by a step-like
potential. This suggests a possible device application of a
nanoswitch based on such a metallic graphene junction.
\par
  This work was financially supported by the National Nature Science
Foundation of China under Grant NNSFC10774055.

\pagebreak
\renewcommand{\baselinestretch}{3}

\clearpage
%\section{\protect\bigskip\ {\protect\large FIGURES}}

\begin{figure}
\centering\scalebox{0.4}{\includegraphics{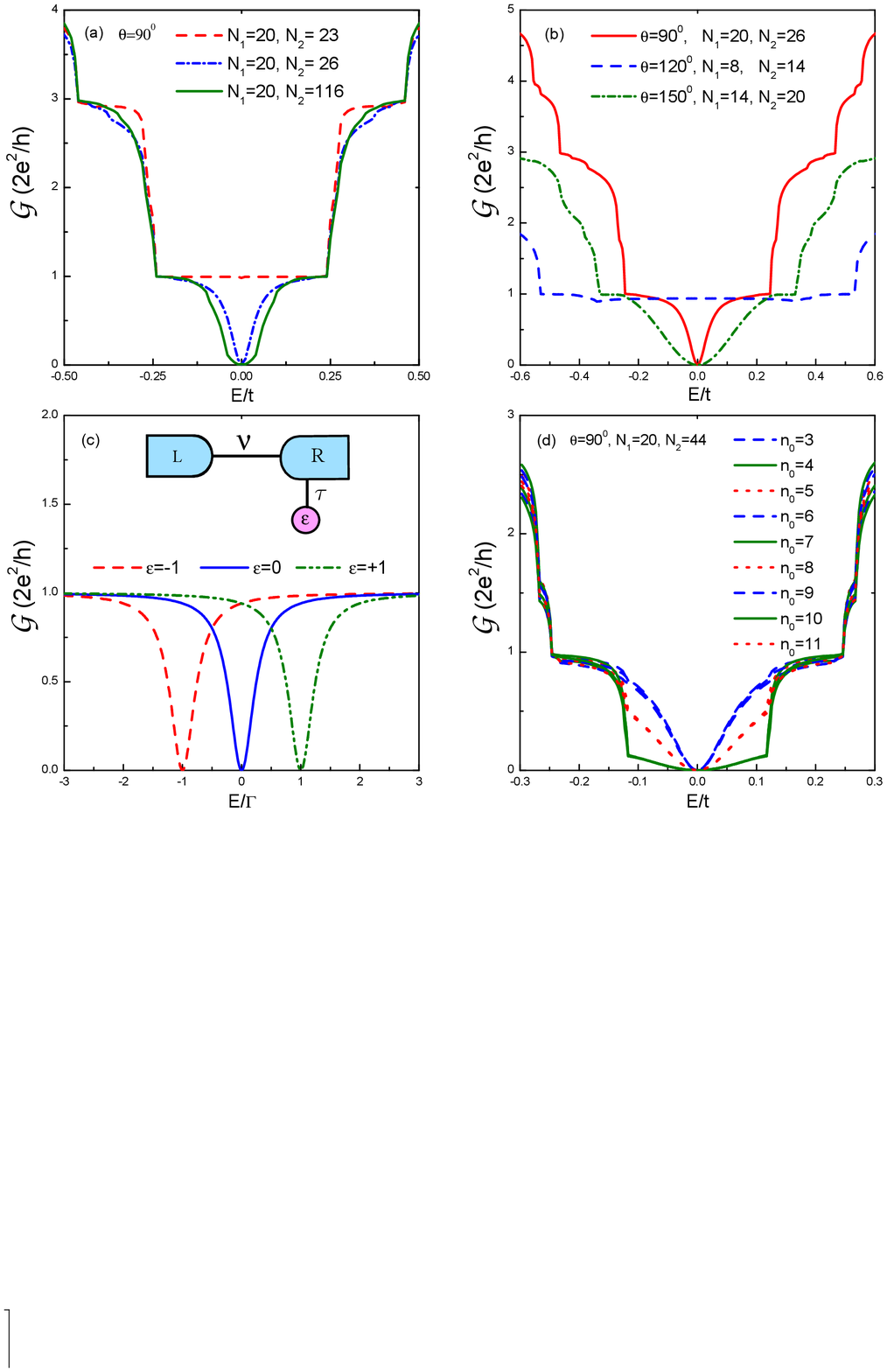}}
\caption{(Color online) (a) The conductance spectrum of some square
junctions, i.e. $\theta=90^{0}$. (b) The conductance spectrum for
the obtuse angle bent junctions, $\theta=120^\circ$ and $150^\circ$.
(c) Schematic of a laterally coupled quantum dot structure, where
$v$ and $\tau$ represent the coupling coefficients between the two
leads and between the quantum dot and the right lead. The calculated
conductance spectrum from this model is also plotted. (d) The
conductance spectra of the graphene junctions with different shapes
created by shifting downwards the narrower ribbon.
\label{conductance1}}
\end{figure}

\begin{figure}
\centering \scalebox{0.4}{\includegraphics{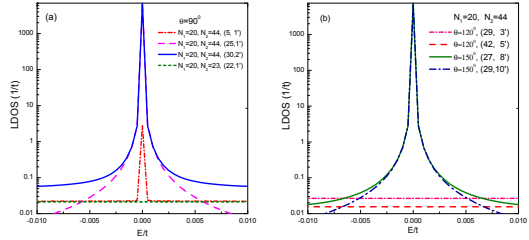}}
\caption{(Color online) The spectrum of the local density of states
at different lattice points near the junction interface of the
typical junctions (a) $\theta=90^\circ$. (b) $\theta=120^\circ$ and
$\theta=150^\circ$. A lattice point is determined by a pair of
indexes (n,n') where n and n' are labeled in Fig.1(a). \label{dos}}
\end{figure}

\begin{figure}
\centering \scalebox{0.4}{\includegraphics{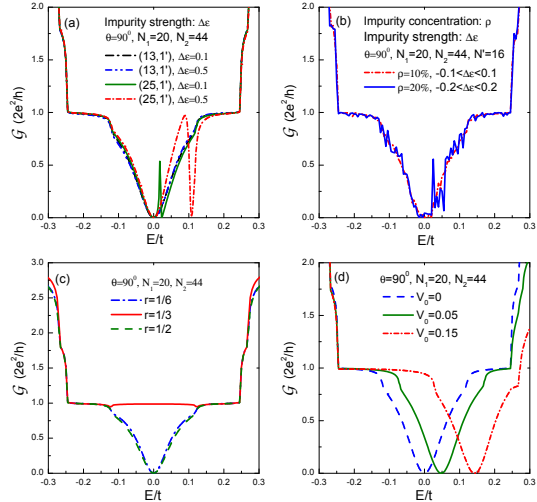}}
\caption{(Color online) The conductance spectrum influenced by some
kinds of scattering. (a) A single impurity positioned at different
lattice points. (b) Many impurities with fluctuating strengths
randomly distributed in the device region. (c) Vacancies appearing
at the zigzag edge of the junction shoulder. (d) A step-like
potential applied to the wider ribbon.  \label{QD}}
\end{figure}
\bigskip
\end{document}